\documentclass[12pt]{JHEP3} 

\usepackage{epsfig,multicol}
\usepackage{amsfonts}
\usepackage{amsmath}

\title{Monodromy Matrix in the PP-Wave Limit}
\author{ Ashok Das$^a$, Jnanadeva Maharana$^{b}$ and A. Melikyan$^a$ \\
\\
$^a$ Department of Physics and Astronomy\\
University of Rochester\\
Rochester, NY 14627-0171, USA\\
\\
$^b$ Institute of Physics\\
Bhubaneswar-751005, India}
\maketitle

\abstract{
We construct the monodromy matrix for a class of gauged WZWN models in the
plane wave limit and discuss various properties of such systems.}

\begin{document}

\newpage

\section{Introduction}

It is recognized that two dimensional field theories have played an important
role in describing a variety of physical systems \cite{das, abd, cft}.
Furthermore, some of these models possess the interesting property of
integrability which corresponds to exactly solvable models. The nonlinear $%
\sigma$-models in $1+1$ dimensions have been studied in great detail from
different perspectives. Such models are endowed with a rich symmetry
structure. The integrability properties of $\sigma$-models
in flat space have attracted considerable attention in the past. 
In the context of Einstein's gravity and supergravity
theories, when a four dimensional action is dimensionally reduced to two
space-time dimensions, the resulting action, in general, describes a $\sigma$
-model on a coset, $G$/$H$, where $G$ turns out to be a noncompact group in
many cases and $H$ is its maximal compact subgroup. In order to study
integrability properties of such models, one of the approaches
involves  constructing 
the monodromy matrix associated with the system under consideration \cite
{bz, nic1, nic2, nic3}. In the case of $\sigma$-models in flat space, it is
customary to introduce a constant spectral parameter and then construct a
suitable current which is curvature free. This single condition 
fulfills the requirement of integrability as well as yields the field
equations. 
On the other hand, when we consider a $\sigma$-model in a curved background,
the spectral parameter assumes space-time dependence to satisfy consistency
requirements in order that the curvaturelessness of the appropriate currents
are maintained.

In the frame work of string theory, we encounter systems with a large number
of isometries. Some of these systems are described by a two dimensional
effective action which may be derived from the dimensional reduction of a ten
dimensional string effective action \cite{ib, jmi, jhs, sen, brunelli}. 
For example, a four dimensional stringy
spherically symmetric black hole can be described by a two dimensional
effective action and such an action is endowed with a rich duality symmetry
group. In addition to the duality symmetry, the theory, in several cases,
possesses additional symmetries special to two dimensional models.

In the past, the symmetry contents of theories, when the fields depend 
only on one of the light-cone (LC) variables, have been explored. One
starts from a 
two dimensional action whose integrability properties are best described
when the equations of motion and integrability conditions (more
appropriately curvaturelessness of the currents with a spectral parameter)
are expressed in terms of the LC variables. One expects an enhancement of
symmetries when we have a theory described by only one coordinate.
Julia was first to propose that such symmetry enhancements might occur
in dimensionally reduced theory where fields depend on a single variable
\cite{julia1}. It is
worthwhile to note that symmetry properties of supergravity
theories/M-theory have been studied in the vicinity of space-like
singularities. Near the singularity, the dynamical evolution of fields at
each spatial point is asymptotically governed by a set of second order
differential equations in time. In other words, the spatial points decouple.
An elegant description can be given in terms of a nonlinear $\sigma$-model in
one dimension and the Kac-Moody algebra makes an appearance \cite{damour1}.
 Thus, it is worthwhile to consider an exactly solvable model in 
order to study some of these features and we 
study the WZWN model at hand in the Penrose limit.

The AdS/CFT duality in string theories \cite{ADSCFT} is a very powerful
result which exhibits the intimate connection between string theory
and supersymmetric Yang-Mills theory.
When one  considers the pp-wave limit of string theory in $AdS_5 \times S_5$
background in the presence of RR flux, the exact spectrum of the
theory can be deduced. Therefore, such a theory may be identified
as an integrable system  
\cite{pp1,pp2,pp3}. Furthermore, in this limit,
 the BMN conjecture \cite{BMN} has been used as a powerful tool to explore
more interesting connections between the string states and the states of
 the $N=4$ supersymmetric
Yang-Mills theory.

The integrability of dimensionally reduced gravity and
supergravity to two dimensions have been studied extensively. As mentioned
earlier, the first step in this direction is to introduce the spectral
parameter and construct a set of currents which satisfy curvaturelessness
condition. In the context of string theory, the graviton and the
shifted dilaton 
appear together in the two dimensional string effective action in 
many problems of interest. The construction of the monodromy matrix and
its transformation properties under T-duality has been investigated
in the recent past \cite{DMM}%

The purpose of this article is to construct the monodromy matrix for
a WZWN model and examine its behavior in the pp-wave limit. The WZWN
models are exactly solvable theories. In the past, we have taken
advantage of this attribute of WZWN models to verify our proposal
regarding duality properties of the monodromy matrix. We find that
the monodromy matrix has interesting behavior in the pp-wave limit.

\section{The General Construction}

In this section, we briefly review the general procedure for
construction of the
monodromy matrix for two dimensional string effective actions. We shall
follow the notations and conventions of our earlier works \cite{DMM}. The
essential point is that the action describes a $\sigma$-model coupled to
gravity in two dimensions.  In the conformal gauge ($g_{\alpha\beta} =
\eta_{\alpha\beta} e^{\bar{\phi}}$), the action for such a sigma model
(derived from dimensional reduction of the effective string theory to two
dimensions) takes the following form \cite{jmj,hasan} 
\begin{equation}
S_{\sigma} = \frac{1}{8} \int \mathrm{d}x^{0}\mathrm{d}x^{1}\, e^{-\bar{\phi}%
} \eta^{\alpha\beta} \mathrm{Tr}\left(\partial_{\alpha} M^{-1}
\partial_{\beta} M\right),  \label{action}
\end{equation}
where $\alpha,\beta = 0,1$, $\bar{\phi}$ the shifted dilaton and the matrix $%
M$ belongs to the T-duality group and we identify it to be  $O(d,d)$, $d$
being the number of isometries when dimensional reduction is implemented.
The matrix M is 
\begin{equation}
M = \left(%
\begin{array}{cc}
G^{-1} & - G^{-1}B \\ 
BG^{-1} & G - BG^{-1}B%
\end{array}%
\right),  \label{M}
\end{equation}
where $G_{ij}$ and $B_{ij}$ represent respectively the metric in the
internal space and the moduli coming from dimensional reduction of the NS-NS
two form potential in higher space-time dimensions. The moduli $G$ and $B$
parameterize the coset ${O(d,d)}$/${O(d) \times O(d)}$. Under the global $O
(d,d)$ transformation, 
\begin{equation}
M \rightarrow \Omega^{T} M \Omega,\qquad \Omega\in O (d,d),
\end{equation}
while the shifted dilaton remains unchanged. The matrix $M$ can be written
in the factorized form 
\begin{equation}
M = V V^{T},\qquad V\in \frac{O(d,d)}{O(d)\times O(d)},
\end{equation}
where $V$ has the triangular form 
\begin{equation}
V = \left(%
\begin{array}{cc}
E^{-1} & 0 \\ 
BE^{-1} & E^{T}%
\end{array}%
\right),  \label{Vmatrix}
\end{equation}
with $\left(E^{T}E\right)_{ij} = G_{ij}$. The matrix $V$ parameterizing the
coset $\frac{O(d,d)}{O(d)\times O(d)}$ transforms nontrivially under global $%
O(d,d)$ as well as a local $O(d)\times O(d)$  as 
\begin{equation}
V\rightarrow \Omega^{T} V h (x),\qquad \Omega\in O(d,d),\quad h (x)\in
O(d,d)\times O(d,d),
\end{equation}
and consequently 
\begin{equation}
M = V V^{T} \rightarrow \Omega^{T} VV^{T} \Omega = \Omega^{T} M \Omega.
\end{equation}

Next, we construct the current $V^{-1}\partial _{\alpha}V $ which belongs
to the Lie algebra of $O(d,d)$ and it can be decomposed as 
\begin{equation}
V^{-1}\partial _{\alpha} V = P_{\alpha} + Q_{\alpha}.  \label{current}
\end{equation}
Here, $Q_{\alpha}$ belongs to the Lie algebra of the maximally compact
subgroup $O(d)\times O(d)$ and $P_{\alpha}$ belongs to the complement.
Furthermore, it follows from the symmetric space automorphism property of
the coset ${\frac{O(d,d)}{O(d)\times O(d)}}$ that $P_{\alpha}^{T} =
P_{\alpha}, Q_{\alpha}^{T} = - Q_{\alpha}$; therefore, 
\begin{eqnarray}
P_{\alpha} & = & {\frac{1}{2}}\left( V^{-1}\partial_{\alpha} V +
(V^{-1}\partial_{\alpha} V)^{T}\right),  \notag \\
Q_{\alpha} & = &{\frac{1}{2}}\left(V^{-1}\partial_{\alpha} V -
(V^{-1}\partial_{\alpha} V)^{T}\right).  \label{curr1}
\end{eqnarray}

It is now straightforward to show that 
\begin{equation}
\mathrm{Tr}\,\left(\partial_{\alpha}M^{-1}\partial_{\beta}M\right) = - 4 
\mathrm{Tr}\,\left( P_{\alpha}P_{\beta}\right).  \label{connection}
\end{equation}
Furthermore, the currents in (\ref{current}) are invariant, under global $%
O(d,d)$ rotations whereas under a local $O(d)\times O(d)$ transformation, $%
V\longrightarrow Vh(x)$, 
\begin{equation}
P_{\alpha}\longrightarrow h^{-1}(x)P_{\alpha}h(x),\qquad Q_{\alpha}
\longrightarrow h^{-1}(x)Q_{\alpha}h(x)+h^{-1}(x)\partial_{\alpha}h(x).
\label{PQ-transform}
\end{equation}
Thus, $Q_{\alpha}$ transforms like a gauge field under local $O(d)\times O(d)
$ transformations, while $P_{\alpha}$ transforms in the adjoint
representation 
under a gauge transformation. It is clear, therefore, that
(\ref{connection}) is invariant under the global $O(d,d)$ as well as
the local $O(d)\times O(d)$ transformations. Consequently, the action
in (\ref{action}) is also 
invariant under the local $O(d)\times O(d)$ transformations.

Let us introduce a one parameter family of matrices $\hat{V}(x,t)$  $t$
being the spectral parameter (we denote time coordinate as $x^0$), such that 
$\hat{V}(x,t=0)=V(x)$ and 
\begin{equation}
\hat{V}^{-1}\partial_{\alpha}\hat{V}=Q_{\alpha}+{\frac{1+t^{2}}{1-t^{2}}}
\,P_{\alpha}+{\frac{2t}{1-t^{2}}}\,\epsilon _{\alpha \beta}P^{\beta}.
\label{x3}
\end{equation}
In the case of a sigma model in the flat space, the spectral parameter
turns out to be a constant. However, when we consider a sigma model in
curved space-time, it is necessary for the spectral parameter to be a
local  function satisfying the first order differential equation 
\begin{equation}
\partial_{\alpha} e^{-\bar{\phi}} = - {\frac{1}{2}}\epsilon _{\alpha
\beta}\partial^{\beta}\left( e^{-\bar{\phi}}\left(t+{\frac{1}{t}}%
\right)\right),  \label{j4}
\end{equation}
in order to fulfill consistency conditions arising from integrability
properties. The solution for the shifted dilaton can be written as a sum
\begin{equation}
\rho (x)=e^{-\bar{\phi} (x)} = \rho_{+} (x^{+}) + \rho_{-} (x^{-}),
\label{j5}
\end{equation}
in terms of which the solution to Eq. (\ref{j4}) is expressed as 
\begin{equation}
t(x)={\frac{\sqrt{\omega +\rho _{+}}-\sqrt{\omega -\rho _{-}}}{\sqrt{\omega
+\rho _{+}}+\sqrt{\omega -\rho _{-}}}},  \label{spectpar}
\end{equation}
where $\omega $ is the constant of integration, which can be thought of as a
global spectral parameter. It is straightforward to check that the zero
curvature condition following from (\ref{x3}) leads to the integrability of
the current in (\ref{current}) as well as the dynamical equation for the
nonlinear sigma model derived from (\ref{action}).

Several comments are in order at this stage which follow  from the
foregoing discussions. First, the one parameter family of connections
(currents) does not determine the potential $\hat{V}(x,t)$ uniquely, namely, 
$\hat{V}$ and $S(\omega )\hat{V}$ , where $S(\omega )$ is a constant matrix,
yield the same one parameter family of connections. Second, in the presence
of the spectral parameter, the symmetric space automorphism can be
generalized as 
\begin{equation}
\eta ^{\infty}(\hat{V}(x,t)) = \eta \left(\hat{V}\left(x,{\frac{1}{t}}%
\right)\right) = \left(\hat{V} ^{-1}\left(x,{\frac{1}{t}}\right)\right)^{T}.
\label{x6}
\end{equation}
Thus the family of matrices, ${\hat V}(x,t)$, satisfy 
\begin{equation}
\left(\hat{V}^{-1}\left(x,{\frac{1}{t}}\right)\partial_{\alpha}\hat{V}
\left(x,{\frac{1}{t}}\right)\right)^{T} = -\hat{V}^{-1}(x,t)\partial_{\alpha}%
\hat{V}(x,t).  \label{x7}
\end{equation}
We are now in a position to define the monodromy matrix as 
\begin{equation}
\mathcal{M}=\hat{V}(x,t)\hat{V}^{T}\left(x,{\frac{1}{t}}\right).  \label{x8}
\end{equation}
It follows, from Eq. (\ref{x7}), that 
\begin{equation}
\partial_{\alpha}\mathcal{M}=0.  \label{x9}
\end{equation}
In other words, $\mathcal{M}=\mathcal{M}(\omega)$ is independent of
space-time coordinates. The monodromy matrix for a system under
investigation encodes integrability properties such as the conserved
charges associated with the theory.

\section{A Wess-Zumino-Witten Model}

We are interested in investigating attributes of the monodromy matrix in the
pp-wave limit. We focus our attention on  a specific WZWN model due to
Sfetsos and Tseytlin \cite{ST}. The pp-wave limit exists for this model. The
model describes a gauged WZWN model based on $\left(E_{2}^{c}\otimes U
(1)\right)/U (1)$, where $E_{2}^{c}$ represents the two dimensional
Euclidean group with a central extension. Following the notation of \cite{ST}%
,  the action is given by 
\begin{eqnarray}
S & = & I_{0}(g_{1})+\frac{1}{\pi}\int d^{2}z\Big[\frac{1}{2}\partial \phi 
\overline{\partial}\phi +iA\left( \sqrt{q_{0}}\overline{\partial}\phi + 
\overline{\partial}x_{1}+\cos u\overline{\partial}x_{2}\right)  \notag \\
\noalign{\vskip 4pt}& &-i\overline{A}\left( \sqrt{q_{0}}\partial \phi
+\partial x_{2}+\cos u\partial x+A\overline{A}(1+\cos u+2q_{0})\right)\Big],
\label{tseytlin}
\end{eqnarray}
where we have identified $g_{1} = E_{2}^{c}, g_{2} = U(1)$ and $I_{0} (g_{1})
$ represents the action for the WZWN model for the group $E_{2}^{c}$. The
group elements are 
\begin{eqnarray}
g & = & \mathrm{diag}\,(g_{1},g_{2}),  \notag \\
\noalign{\vskip 4pt}g_{1} & = &
e^{ix_{1}P_{1}}e^{iuJ}e^{ix_{2}P_{2}}e^{ivF}\in E_{2}^{c},  \notag \\
\noalign{\vskip 4pt}g_{2} & = & e^{i\phi P_{0}}\in U(1).
\end{eqnarray}
Here, $P_{i}, J, i=1,2$ represent the generators of the two dimensional
Euclidean group, $F$ the central extension generator and $P_{0}$ is the
generator for $U (1)$. Note that $q_{0}$ appearing in (\ref{tseytlin}) is a
constant parameter which parameterizes the embedding of $U(1)$ into $%
E_{2}^{c}\otimes U(1)$.

The action (\ref{tseytlin}) is invariant under the infinitesimal gauge
transformations 
\begin{equation}
\delta x_{1} = \delta x_{2} = \alpha (x),\quad \delta \phi = 2\sqrt{q_{0}}
\alpha (x),\quad \delta A = i\partial\alpha (x),\quad \delta \bar{A} = - i 
\bar{\partial}\alpha (x),
\end{equation}
Furthermore, $u$ and $v$ remain inert under the gauge transformation. Thus
we use this property to fix the gauge $\phi = 0$.  The metric, the
anti-symmetric tensor fields as well as the shifted dilaton can be
identified from (\ref{tseytlin}) (after some rescaling) to be 
\begin{eqnarray}
\mathrm{d}s^{2} & = & 2\mathrm{d}v\mathrm{d}u+\frac{\cos
  ^{2}u}{q_{0}+\cos ^{2}u}\mathrm{d}x_{1}^{2}+\frac{\sin
^{2}u}{q_{0}+\cos ^{2}u}\mathrm{d}x_{2}^{2},  \label{1} \\
\noalign{\vskip 4pt}B_{12} & = & - B_{21} = b = \frac{\sqrt{q_{0}(1+q_{0})}}{%
q_{0}+\cos ^{2}u},  \label{2} \\
\noalign{\vskip 4pt}\bar{\phi} & = & - \ln \sin^{2} 2u,  \label{2'}
\end{eqnarray}
where we have chosen a specific value for the constant (in the expression
for the dilaton) for later convenience. We note that the anti-symmetric
tensor field $b = 0$ when $q_{0}=0$. We recall, from our earlier
experience, that it might be convenient to start from the following set of
background configurations 
\begin{eqnarray}
\mathrm{d}s^{2} & = & 2\mathrm{d}v\mathrm{d}u+\mathrm{d}x_{1}^{2}+
\tan^{2} u\, \mathrm{d}x_{2}^{2},  \label{3} \\
\noalign{\vskip 4pt}b_{0} & = & 0,  \label{4} \\
\noalign{\vskip 4pt}\bar{\phi}_{0} & = & - \ln \sin^{2} 2u,  \label{4'}
\end{eqnarray}
and then implement an appropriate T-duality transformation to obtain
the  backgrounds
associated with the action (\ref{tseytlin}).

Let us note from the structure of the backgrounds in (\ref{tseytlin})
as well as from the definition of the $M$-matrix 
\begin{equation}
M=VV^{T}=\left( 
\begin{array}{cc}
G^{-1} & -G^{-1}B \\ 
BG^{-1} & G-BG^{-1}B%
\end{array}
\right),  \label{5}
\end{equation}
that we can identify 
\begin{equation}
G = \left(%
\begin{array}{cc}
G_{11} & 0 \\ 
0 & G_{22}%
\end{array}%
\right),
\end{equation}
where
\begin{equation}
G_{11} = \frac{\cos^{2} u}{q_{0} + \cos^{2} u},
\quad G_{22} = \frac{\sin^{2} u}{q_{0} + \cos^{2} u} ,  \label{5'}
\end{equation}
and the anti-symmetric field is given by (\ref{2}). For $q_{0}=0$, on the
other hand, we have (see (\ref{3}), (\ref{4}) and (\ref{5})) 
\begin{equation}
M_{0} = \left( 
\begin{array}{cc}
G_{0}^{-1} & 0 \\ 
0 & G_{0}%
\end{array}
\right),  \label{6a}
\end{equation}
with 
\begin{equation}
G_{0} = \left( 
\begin{array}{cc}
1 & 0 \\ 
0 & \tan^{2} u%
\end{array}%
\right).  \label{6bb}
\end{equation}
It is straightforward to show from these that the two matrices, $M$ and $%
M_{0}$, are related through 
\begin{equation}
M = \Omega^{T} M_{0} \Omega,  \label{rotation}
\end{equation}
where the rotation matrix $\Omega$ has the explicit form 
\begin{equation}
\Omega =\left( 
\begin{array}{cccc}
\sqrt{1+q_{0}} & 0 & 0 & -\sqrt{q_{0}} \\ 
0 & \sqrt{1+q_{0}} & \sqrt{q_{0}} & 0 \\ 
0 & \sqrt{q_{0}} & \sqrt{1+q_{0}} & 0 \\ 
-\sqrt{q_{0}} & 0 & 0 & \sqrt{1+q_{0}}%
\end{array}
\right).  \label{7}
\end{equation}

\section{The Monodromy matrix}

The construction of the monodromy matrix for this model can be carried
out in a manner parallel to the case of the Nappi-Witten model
\cite{NW}. For $q_{0}=0$, we can write (see (\ref{Vmatrix}) and (\ref{6bb})) 
\begin{equation}
V^{(B=0)} = \left(%
\begin{array}{cc}
\left(E^{(B=0)}\right)^{-1} & 0 \\ 
0 & E^{(B=0)}%
\end{array}%
\right),\qquad E^{(B=0)} = \left(%
\begin{array}{cc}
1 & 0 \\ 
0 & \tan u%
\end{array}%
\right),  \label{V}
\end{equation}
so that we can determine (see (\ref{curr1}) and we will use light-cone
variables where we identify $u = x^{+}, v= x^{-}$) 
\begin{eqnarray}
P^{(B=0)}_{+} & = & \left(%
\begin{array}{cccc}
0 & 0 & 0 & 0 \\ 
0 & -\frac{4}{\sin 2x^{+}} & 0 & 0 \\ 
0 & 0 & 0 & 0 \\ 
0 & 0 & 0 & \frac{4}{\sin 2 x^{+}}%
\end{array}%
\right),  \notag \\
\noalign{\vskip 4pt}P^{(B=0)}_{-} & = & 0 = Q^{(B=0)}_{\pm}.  \label{P+}
\end{eqnarray}
Therefore, the one parameter family of potentials $\hat{V}^{(B=0)} (x,t)$
satisfy (since $B=0$) 
\begin{equation}
\left(\hat{V}^{(B=0)}\right)^{-1} \partial_{\pm} \hat{V}^{(B=0)} = \frac{%
1\mp t}{1\pm t}\, P_{\pm},  \label{Vhat}
\end{equation}
where $t$ denotes the space-time dependent spectral parameter.

From the form of $P_{+}$ in (\ref{P+}), we expect $\hat{V}^{(B=0)}$ in (\ref%
{Vhat}) to be diagonal. We note that, with the identification of the shifted
dilaton in (\ref{2'}) (or (\ref{4'})), we obtain the solution 
\begin{equation}
\rho = e^{-\bar{\phi}} = \rho_{+} (x^{+}) + \rho (x^{-}),\quad \rho_{+} = 
\frac{1}{2} \left(1-2 \cos^{2} 2x^{+}\right),\quad \rho_{-} = \frac{1}{2}.
\label{rho}
\end{equation}
In this case, the spectral parameter does not depend on $x^{-}$ and
satisfies the equation (see (\ref{j4})) 
\begin{equation}
\partial_{+} t = - \frac{4t (1-t)}{(1+t)}\, \frac{\cos 2x^{+}}{\sin 2x^{+}}.
\end{equation}
Following our earlier method \cite{DMM}, we can now determine the diagonal
form of the matrix $\hat{V}$ to be 
\begin{eqnarray}
\hat{V}^{(B=0)} (x, t) & = & \mathrm{diag}\, \left(\hat{V}_{1}, \hat{V}_{2}, 
\hat{V}_{3}, \hat{V}_{4}\right)  \notag \\
\noalign{\vskip 4pt}& = & \mathrm{diag}\left(1,~ \frac{(1-t) \tan x^{+}}{t+
\tan^{2} x^{+}},~ 1, ~ \frac{t+ \tan^{2} x^{+}}{(1-t) \tan x^{+}}\right).
\end{eqnarray}
We note that, for the case at hand, we have $t_{2}=1$ and $t_{4}= - \tan^{2}
x^{+}$ corresponding to $\omega_{2} = \frac{1}{2} = - \omega_{4}$
respectively. It is easy to check that when $t=0$, the matrix $\hat{V}%
^{(B=0)}$ indeed reduces to $V^{(B=0)}$ in (\ref{V}). The monodromy matrix
can be easily constructed now and has the form 
\begin{eqnarray}
\mathcal{M}^{(B=0)} & = & \hat{V}^{(B=0)} (x,t) \left(\hat{V}%
^{(B=0)}\right)^{T} \left(x, \frac{1}{t}\right)  \notag \\
\noalign{\vskip 4pt} & = & \left(%
\begin{array}{cccc}
1 & 0 & 0 & 0 \\ 
\noalign{\vskip 4pt}0 & \frac{1-2\omega}{1+2\omega} & 0 & 0 \\ 
\noalign{\vskip 4pt}0 & 0 & 1 & 0 \\ 
\noalign{\vskip 4pt}0 & 0 & 0 & \frac{1+2\omega}{1-2\omega}%
\end{array}%
\right).  \label{Mcal}
\end{eqnarray}
The poles of the monodromy matrix at $\omega_{2}= \frac{1}{2} = - \omega_{4}$
are now manifest. However, the other two trivial diagonal elements of the
monodromy matrix would seem to suggest that we are missing two other poles.
A careful analysis shows that the location of the other two poles has moved
to infinity. Thus we recognize that for $\omega_{1} = -
\omega_{3}\rightarrow \infty$, we obtain the following relations 
\begin{equation}
\frac{\omega_{1}-\omega}{\omega_{1}+\omega} = 1 = \frac{\omega_{1}+\omega}{%
\omega_{1}-\omega}.
\end{equation}
In other words, in the pp-wave limit, the monodromy matrix for our model has
even a simpler structure than that of a generic WZWN model.

Notice that having constructed the monodromy matrix for the case $b=0$ (or $%
q_{0}=0$), and analyzed its properties, we are in a position to generalize
the results to the case for nontrivial values of the anti-symmetric tensor
field. We note that 
\begin{equation}
V^{(B)} (x) = \left(%
\begin{array}{cc}
\left(E^{(B)}\right)^{-1} & 0 \\ 
\noalign{\vskip 4pt}0 & E^{(B)}%
\end{array}%
\right) = \Omega^{T} V^{(B=0)} (x) h (x),  \label{VB}
\end{equation}
where $\Omega$ represents the global rotation defined in (\ref{7} and $h(x)
\in H$ which is the maximal compact subgroup of the T-duality group. 
\begin{equation}
E^{(B)} = \frac{1}{\sqrt{q_{0}+\cos^{2} x^{+}}}\left(%
\begin{array}{cc}
\cos x^{+} & 0 \\ 
\noalign{\vskip 4pt}0 & \sin x^{+}%
\end{array}%
\right),
\end{equation}
and 
\begin{equation}
h (x) = \left(%
\begin{array}{cccc}
\cos\theta & 0 & 0 & \sin\theta \\ 
\noalign{\vskip 4pt}0 & \cos\theta & -\sin\theta & 0 \\ 
\noalign{\vskip 4pt}0 & \sin\theta & \cos\theta & 0 \\ 
\noalign{\vskip 4pt}-\sin\theta & 0 & 0 & \cos\theta%
\end{array}%
\right),  \label{h}
\end{equation}
where the angular variable $\theta$ is defined by 
\begin{equation}
\sin^{2} \theta = \frac{q_{0}\sin^{2} x^{+}}{q_{0} + \cos^{2} x^{+}}.
\label{theta}
\end{equation}

It follows now directly from the above relations that 
\begin{eqnarray}
P_{+}^{(B)} & = & h^{-1} (x) P_{+}^{(B=0)} h (x) = \frac{2}{q_{0} + \cos^{2}
x^{+}}  \notag \\
\noalign{\vskip 4pt}& &\times\left(%
\begin{array}{cccc}
q_{0}\tan x^{+} & 0 & 0 & -\sqrt{q_{0}(q_{0}+1)} \\ 
\noalign{\vskip 4pt}0 & -\frac{(q_{0}+1)}{\tan x^{+}} & \sqrt{q_{0}(q_{0}+1)}
& 0 \\ 
\noalign{\vskip 4pt}0 & - \sqrt{q_{0}(q_{0}+1)} & - q_{0}\tan x^{+} & 0 \\ 
\noalign{\vskip 4pt}\sqrt{q_{0}(q_{0}+1)} & 0 & 0 & \frac{(q_{0}+1)}{\tan
x^{+}}%
\end{array}%
\right),  \notag \\
\noalign{\vskip 4pt}P_{-}^{(B)} & = & h^{-1} (x) P_{-}^{(B=0)} h (x) = 0, 
\notag \\
\noalign{\vskip 4pt}Q_{+}^{(B)} & = & h^{-1} (x) \partial_{+} h(x) + h^{-1}
(x) Q_{+}^{(B=0)} h (x)  \notag \\
\noalign{\vskip 4pt}& = & \frac{2\sqrt{q_{0}(q_{0}+1)}}{q_{0}+\cos^{2} x^{+}}
\left(%
\begin{array}{cccc}
0 & 0 & 0 & 1 \\ 
\noalign{\vskip 4pt}0 & 0 & -1 & 0 \\ 
\noalign{\vskip 4pt}0 & 1 & 0 & 0 \\ 
\noalign{\vskip 4pt}-1 & 0 & 0 & 0%
\end{array}%
\right),  \notag \\
\noalign{\vskip 4pt}Q_{-}^{(B)} & = & h^{-1} (x)\partial_{-} h (x) + h^{-1}
(x) Q_{-}^{(B=0)} h (x) = 0.
\end{eqnarray}
We have computed $P^{(B)}_{\pm}$ and $Q^{(B)}_{\pm}$ from the ST model
directly and verified that $P^{(B)}_{\pm}$ and $Q^{(B)}_{\pm}$ obtained
through the duality transformations from the $B=0$ model agree with these
direct computations.

The one parameter family of potentials can also be constructed in a
straightforward manner and take the form 
\begin{equation}
\hat{V}^{(B)} (x, t) = \Omega^{T} \hat{V}^{(B=0)} h (x) = \left(%
\begin{array}{cccc}
a_{1} & 0 & 0 & a_{2} \\ 
\noalign{\vskip 4pt}0 & a_{3} & a_{4} & 0 \\ 
\noalign{\vskip 4pt}0 & a_{5} & a_{6} & 0 \\ 
\noalign{\vskip 4pt}a_{7} & 0 & 0 & a_{8}%
\end{array}%
\right),  \label{VhatB}
\end{equation}
with 
\begin{eqnarray}
a_{1} & = & \sqrt{1+q_{0}} \cos\theta + \sqrt{q_{0}} \hat{V}_{4} \sin\theta,
\\
\noalign{\vskip 4pt}a_{2} & = & \sqrt{1+q_{0}} \sin\theta - \sqrt{q_{0}} 
\hat{V}_{4} \cos\theta, \\
\noalign{\vskip 4pt}a_{3} & = & \sqrt{q_{0}} \sin\theta + \sqrt{1+q_{0}} 
\hat{V}_{2} \cos\theta, \\
\noalign{\vskip 4pt}a_{4} & = & \sqrt{q_{0}} \cos\theta - \sqrt{1+q_{0}} 
\hat{V}_{2} \sin\theta, \\
\noalign{\vskip 4pt}a_{5} & = & \sqrt{1+q_{0}} \sin\theta + \sqrt{q_{0}} 
\hat{V}_{2} \cos\theta, \\
\noalign{\vskip 4pt}a_{6} & = & \sqrt{1+q_{0}} \cos\theta - \sqrt{q_{0}} 
\hat{V}_{2} \sin\theta, \\
\noalign{\vskip 4pt}a_{7} & = & -\sqrt{q_{0}} \cos\theta + \sqrt{1+q_{0}} 
\hat{V}_{4} \sin\theta, \\
\noalign{\vskip 4pt}a_{8} & = & -\sqrt{q_{0}} \sin\theta + \sqrt{1+q_{0}} 
\hat{V}_{4} \cos\theta,
\end{eqnarray}
with $\hat{V}_{i}$ and $\theta$ defined respectively in (\ref{Vhat}) and (%
\ref{theta}). In turn, this gives us the monodromy matrix for the general
case of the form 
\begin{eqnarray}
\mathcal{M}^{(B)} & = & \hat{V}^{(B)} (x,t) \left(\hat{V}^{(B)}\right)^{T}
\left(x, \frac{1}{t}\right)  \notag \\
\noalign{\vskip 4pt} & = & \left(%
\begin{array}{cccc}
b_{1} & 0 & 0 & b_{2} \\ 
\noalign{\vskip 4pt}0 & b_{3} & b_{4} & 0 \\ 
\noalign{\vskip 4pt}0 & b_{4} & b_{5} & 0 \\ 
\noalign{\vskip 4pt}b_{2} & 0 & 0 & b_{6}%
\end{array}%
\right),
\end{eqnarray}
where 
\begin{eqnarray}
b_{1} & = & 1 + \frac{2q_{0}}{1-2\omega}, \\
\noalign{\vskip 4pt}b_{2} & = & -\sqrt{q_{0}(1+q_{0})} - \sqrt{q_{0}} \frac{%
1+2\omega}{1-2\omega}, \\
\noalign{\vskip 4pt}b_{3} & = & -1 + \frac{2(1+q_{0})}{1+2\omega}, \\
\noalign{\vskip 4pt}b_{4} & = & \sqrt{q_{0}} + \sqrt{q_{0}(1+q_{0})} \frac{%
1-2\omega}{1+2\omega}, \\
\noalign{\vskip 4pt}b_{5} & = & 1 + \frac{2q_{0}}{1+2\omega}, \\
\noalign{\vskip 4pt}b_{6} & = & -1 + \frac{2(1+q_{0})}{1-2\omega}.
\end{eqnarray}
There are several interesting features to be noted from the form of the
general monodromy matrix. First, it is straightforward to check that 
\begin{equation}
\mathcal{M}^{(B)} = \Omega^{T} \mathcal{M}^{(B=0)} \Omega
\end{equation}
and that it reduces to the monodromy matrix (\ref{Mcal}) for the simpler
background i.e. for vanishing anti-symmetric field when $q_{0}=0$.

The monodromy matrix for the Nappi-Witten model can also be obtained in the
pp-wave limit adapting the same procedure. We may recall that  in the case
of the Nappi-Witten model 
\begin{eqnarray}
\mathrm{d} s^{2} & = & - \mathrm{d}\tau^{2} + \mathrm{d} x^{2} + \frac{1}{%
1-\cos 2\tau \cos 2x}  \notag \\
\noalign{\vskip 4pt} & & \times 4\left(\cos^{2} \tau \cos^{2} x \mathrm{d}%
y^{2} + \sin^{2} \tau \sin^{2} x \mathrm{d}z^{2}\right),  \notag \\
\noalign{\vskip 4pt}\phi & = & - \frac{1}{2}\,\ln \left(1-\cos 2\tau \cos
2x\right),  \notag \\
\noalign{\vskip 4pt}B_{12} & = & - B_{21} = b = \frac{(\cos 2\tau - \cos 2x)%
}{(1-\cos 2\tau \cos 2x)}.  \label{NW}
\end{eqnarray}
Here $\tau$ represents time (to avoid confusion with the spectral parameter)
and we have set an arbitrary constant parameter of the model to zero for
conveniences.

Let us define the light-cone coordinates 
\begin{equation}
x^{+} = \frac{\tau + x}{\sqrt{2}},\qquad x^{-} = \frac{\tau - x}{\sqrt{2}},
\end{equation}
and take the pp-wave limit $v\rightarrow 0$. Note from (\ref{NW}) that the
anti-symmetric tensor field vanishes in this limit and we have 
\begin{eqnarray}
\mathrm{d} s^{2} & = & - 2 \mathrm{d} x^{+} \mathrm{d} x^{-} + \tan^{-2} 
\frac{x^{+}}{\sqrt{2}} \mathrm{d} y^{2} + \tan^{2} \frac{x^{-}}{\sqrt{2}} 
\mathrm{d} z^{2},  \notag \\
\noalign{\vskip 4pt}\bar{\phi} & = & - \ln \sin^{2} \sqrt{2}x^{+},
\end{eqnarray}
where $\bar{\phi}$ represents the shifted dilaton as before. In this case,
the one parameter family of connections can be obtained. For the  specific
identifications $\omega_{1} = \omega_{2} = \frac{1}{2}$ we have two
independent solutions (depending on the choice of the sign  of the square
root) and we get 
\begin{equation}
\hat{V} (x, t) = \mathrm{diag}\,\left(\hat{V}_{1}, \hat{V}_{2}, \hat{V}_{3}, 
\hat{V}_{4}\right),
\end{equation}
where 
\begin{eqnarray}
\hat{V}_{1} & = & - \frac{t-1}{t \tan^{2} \frac{x^{+}}{\sqrt{2}} + 1} = \hat{%
V}_{3}^{-1},  \notag \\
\noalign{\vskip 4pt}\hat{V}_{2} & = & - \frac{(t-1)\tan^{2} \frac{x^{+}}{%
\sqrt{2}}}{t - \tan^{2} \frac{x^{+}}{\sqrt{2}}} = \hat{V}_{4}^{-1}.
\end{eqnarray}
We note that when $t=0$, this reduces to $V$-matrix for this theory. The
monodromy matrix is now given by 
\begin{eqnarray}
\mathcal{M}^{(NW)} & = & \hat{V} (x, t) \left(\hat{V}\right)^{T} \left(x, 
\frac{1}{t}\right)  \notag \\
\noalign{\vskip 4pt} & = & \left(%
\begin{array}{cccc}
\frac{1-2\omega}{1+2\omega} & 0 & 0 & 0 \\ 
\noalign{\vskip 4pt}0 & \frac{1-2\omega}{1+2\omega} & 0 & 0 \\ 
\noalign{\vskip 4pt}0 & 0 & \frac{1+2\omega}{1-2\omega} & 0 \\ 
\noalign{\vskip 4pt}0 & 0 & 0 & \frac{1+2\omega}{1-2\omega}%
\end{array}%
\right).
\end{eqnarray}
Therefore, we conclude that in the pp-wave limit the monodromy matrix in the
Nappi-Witten model has degenerate poles.

\section{Summary and Discussion}

In this paper we have constructed the monodromy matrix for a one dimensional 
$\sigma $-model arising in the pp-wave limit of the gauged WZWN
$\sigma$-model \cite{ST}. In order to study the integrability
properties of the 
effective action, we introduced the spectral parameter, $t$,
following the standard procedure. It has been argued that $t$ is space-time
dependent from consistency requirements and the integration constant,
$\omega $, may also be identified as another global spectral
parameter. Furthermore,
as we have shown, it is possible to trade $\omega $ for $t$ in constructing
the monodromy matrix. Let us recall that with the introduction of $t$ 
(alternatively $\omega $), we are led to envisage a `vielbein', ${\hat{E}}(x,t)
$, which depends on the continuous parameter $t$ besides space-time
coordinates $x$. When we set $t=0$, we get back $E(x)$ which is utilized to
define the action and in turn the action is invariant under the global symmetry
group $G$ and the local group $H$. In defining $\hat{E}$, we have introduced an
infinite family of matrices parameterized by a continuous spectral parameter.
Indeed, the appearance of the infinite-dimensional symmetries of Kac-Moody
type, arising in such two dimensional integrable models, can be
intuitively understood from this perspective. It has been argued \cite
{nic2, julia1} that this symmetry can be even larger when the theory is
reduced to one dimension. In the present case, however, the algebra of
charges can be worked out in a manner completely parallel to that in
\cite{nic2} and we find that it coincides with the algebra in
\cite{nic2}. We do not find any further enhancement of the
algebra beyond what is already known under similar situation. In
several known cases, the resulting symmetry is
of hyperbolic type \cite{kac1}. For instance, the hyperbolic algebra $E_{10}$ 
\cite{damour1, wakimoto, nicolai2, alexk} has been argued to be the
symmetry of the eleven-dimensional
supergravity dimensionally reduced to one dimension \cite{julia1,
  mizoguchi}. Furthermore, when four-dimensional supergravity is
reduced to one dimension, the
resulting symmetry has been shown to be the hyperbolic extension of
$A_{1}^{1}$  algebra
\cite{nicolai1}. However, these hyperbolic algebras do not generally allow
supersymmetric extensions. Therefore, such an enlarged symmetry should
already contain
both bosonic and fermionic generators. From this perspective, it
would be interesting to study the supersymmetric generalization of the $
\sigma $-model considered in this paper. We speculate that, in such a
case, one should
find a hyperbolic extension of an algebra reminiscent of $A_{1}^{1}$. We
refer the reader to the review article \cite{nic2} for a detailed
discussion of these issues and for further references. \newline

\vskip .7cm \noindent \textbf{Acknowledgment:} This work is supported in
part by US DOE Grant No. DE-FG 02-91ER40685.

\end{document}